\documentclass[fleqn, usenatbib, useAMS]{mnras}
\usepackage{newtxtext,newtxmath}
\usepackage{multirow}
\usepackage{multicol}
\usepackage[T1]{fontenc}
\usepackage{ae,aecompl}
\usepackage{pdflscape}
\usepackage{orcidlink}

\usepackage{graphicx}	
\usepackage{amsmath}	
\usepackage{ulem}

\title[Blue Stragglers in M67]{UOCS.VI. UVIT/AstroSat detection of low-mass white dwarf companions to 4 more blue stragglers in M67}

\author[Pandey, Subramaniam and Jadhav]{Sindhu Pandey,$^{1}$\thanks{E-mail: sindhu@aries.res.in, msindhupandey@gmail.com}\orcidlink{0000-0001-6006-1727}
Annapurni Subramaniam,$^{2}$ and Vikrant V. Jadhav,$^{2,3}$\orcidlink{0000-0002-8672-3300}  
\\
$^{1}$Aryabhatta Research Institute of Observational Sciences, Manora Peak, Nainital 263001  India\\
$^{2}$Indian Institute of Astrophysics, Koramangala II Block, Bangalore 560034, India\\
$^{3}$Joint Astronomy Program and Physics Department, Indian Institute of Science, Bangalore 560012 , India\\
}

\date{Accepted XXX. Received YYY; in original form ZZZ}

\pubyear{2021}

\begin{document}

\label{firstpage}
\pagerange{\pageref{firstpage}--\pageref{lastpage}}
\maketitle

\begin{abstract}
Blue stragglers stars (BSSs) in M67 have attracted attention from observations and theory to unravel their formation mechanisms. In the series of Ultra Violet Imaging Telescope (UVIT on \textit{AstroSat}) study of Open clusters (UOCS), here we report the detection of hot companions to 4 more BSSs (WOCS 2013, WOCS 3013, WOCS 4006 and WOCS 5005), using Far-UV photometry obtained in two epochs from UVIT. We characterise the hot companions to be low to extremely low mass (LM/ELM) white dwarfs (WDs) with T$_{eff}$ $\sim$ 13-23kK, R/R$_\odot$ $\sim$ 0.035-0.051, M/M$_\odot$ $\sim$ 0.19-0.3 and age $\sim$ 25-300 Myr using WD models. Two BSSs (WOCS 1025 and WOCS 3005) showed UV excess, and may have a hot companion, but we are unable to confirm/characterise. 13 BSSs are detected by UVIT in the two epochs of data, of the total 14 present in M67. We have performed detailed analysis on 10 BSSs, including our previous studies. Five BSSs are found to have LM/ELM WD companions, suggesting Case-B mass-transfers (MT) to be prevalent in M67 (with a lower limit of 37.5\%, 5/14), along with other mechanisms. Three BS+WD systems have orbital parameters outside the limit for stable MT as per the models. We speculate the following three possibilities - their orbits are altered due to cluster dynamics, some may be in triple systems with LM/ELM WDs in an unknown closer orbit, or a modified MT mechanism may be required to enable their formation. 
\end{abstract}




\section{Introduction}
Evolution in a close binary system often changes the evolutionary pathway of both the stars and is distinctly different from what would happen in isolation \citep{Paczynski1971}. One outcome of such interaction is a blue straggler star (BSS) that prominently lie above the main sequence (MS) turn-off in the cluster colour-magnitude diagram (CMD), a region from which most of the stars of similar mass and age have evolved. They are observed in various stellar environments including open clusters \citep{Ahumada2007}, globular clusters \citep{Krolik1983}, Galactic field \citep{Carney2001,Carney2005,Ekanayake2018} and dwarf spheroidal galaxies \citep{Mapelli2007}, since its first discovery in the globular cluster M3 \citep{1953Sandage}. 
Most of the BSSs are likely to be in a binary system, with a fraction of 76 per cent observed in the old open cluster NGC 188 \citep{Mathieu2009}. The binary origin for BSSs has also been observed in globular clusters \citep{Knigge2009, sahu_2019, Singh2020}. Several theories give possible explanations for the formation of BSSs, with three widely accepted mechanisms in the literature: (i) stellar collision during dynamical encounters in dense environment \citep{Hills1976}, (ii) mass transfer (MT) through Roche lobe overflow in binary systems \citep{McCrea1964} and (iii) mergers or MT of the inner binaries induced by Kozai cycles in hierarchical triple systems \citep{Perets2009, Naoz2014}. While most of the BSSs formed support these models, there exists some BSSs that fail to be explained by these mechanisms \citep{Cannon2015ebss}.

Observational evidence of BSSs with white dwarf (WD) companions have implied MT as a possible route for the formation of these stars \citep{Gosnell2015, Gosnell2019, Sindhu2019, Sindhu2020}. In low-density environments such as open clusters, direct single-single collision is less likely. These stars could be formed through case A, B or C MT, depending on the onset of MT within the MS band, before the core-helium burning stage or after core-helium exhaustion of the donor \citep{Kippenhahn1967, Lauterborn1970}. The consequence of MT would lead to a markedly different phase of the evolution of both stars in the binary system for each case. Case A MT is believed to result in the coalescence of two stars, leading to the formation of single massive BSS or an unmerged system would form a less massive BSS with a short period MS companion. Case B MT produces a BSS from a slightly more evolved companion, which would likely become a short period binary. The donor's remnant would consist of a helium core less than 0.45 M$\odot$ if the initial mass of the donor was less than 3 M$\odot$. As the mass of the remnant is too small to ignite the helium, the star becomes a He core WD. In a Case C MT, the donor is evolved with a large radius and will become a CO WD, with a typical mass of $\ge$ 0.5 M$\odot$. The Case C MT would lead to the formation of BSSs in long-period binaries \citep{Perets2015}.

M67 (NGC 2682) is an old open cluster located at a distance of 850 pc, with an estimated age of 4.2 $\pm$ 0.2 Gyr \citep{Barnes2016}, reddening E(B$-$V) = 0.041 $\pm$ 0.004 mag \citep{Taylor2007} and solar metallicity. The cluster has one of the rich, well studied population of BSSs, with over 25 candidates discussed in literature \citep{Manteiga1989, Gilliland1992, Deng99, Bruntt2007, Lu2010, Geller2015}.
 Based on the proper motion and radial velocity membership probability, \cite{Geller2015} have identified 14 of 25 candidates as members. \cite{Latham1996} have characterised spectroscopic binaries (SB) in M67, prioritising BSSs. Among the 14 BSSs, 10 stars are single-lined spectroscopic binaries (SB1), and one is a double-line spectroscopic binary (SB2). It is observed that most of the BSSs in M67 are in binary orbits, with a high binary frequency of $\sim$79$\pm$24\% \citep{Geller2015}. 
 M67 has been a primary laboratory to study BSSs since its discovery by \cite{Johnson1955}. M67, in fact, has a large number of BSSs for its age and size, and also of diverse nature, which instils interest \citep{Hurley2005}. The N-body model by \cite{Hurley2005} were able to reproduce the cluster properties and closely match the number of BSSs observed in M67. Their successful model produced 20 BSSs with one BS+WD binary. The result of their simulation showed that BSSs in M67 were formed through a variety of pathways, and the dynamics of the cluster environment has a crucial role. \cite{Shetrone2000} carried out spectroscopic observations of 5 BSSs (S984, S975, S1082, S997 and S2204), of which 4 BSSs did not show any signature of composition modification that occurs due to the formation process. 

Ultraviolet (UV) observations have the potential to characterise the formation channels of BSSs and post-MT systems by detecting a possible hot companion to these stars \citep{Landsman1997, Knigge2008, Gosnell2015, Subramaniam2016, sahu_2019, Sindhu2019, Sindhu2020, Jadhav2019, Subramaniam2020}. \cite{Landsman98} were able to detect 11 BSSs in M67 using \textit {Ultraviolet Imaging Telescope (UIT)}. They observed UV excess in two BSSs (S975 and S1082) and suggested the possibility of hot companions. \cite{Siegel2014} studied three open clusters and a globular cluster with Ultraviolet Optical telescope (UVOT) on \textit{Swift}. They studied the BSSs and unusual UV bright stars to demonstrate that UVOT is an excellent instrument to characterise peculiar stars. \cite{Sindhu2018} detected 18 BSSs in the \textit{Galaxy Evolution Explorer} ({\it GALEX}) UV image of M67 and suggested that the BSSs were formed in the last 2.5 Gyr - 400 Myr, nearly continuously. Motivated by these results, we started a survey to understand BS systems in old open clusters using the Ultra Violet Imaging Telescope (UVIT) on \textit{AstroSat} (UVIT Open Cluster Survey - UOCS). UVIT observation of M67 detected a WD companion $<$0.3M$\odot$ to a BSS and several post-MT systems \citep{Sindhu2019, Jadhav2019, Sindhu2020, Subramaniam2020}. 
Here we present the analysis of BSSs in M67 observed with far-UV (FUV) filters of UVIT. We combine the UVIT magnitudes with other estimations in the UV, optical and near-infrared from space as well as ground observations to create multi-wavelength spectral energy distribution (SED).

The paper is structured as follows: The observations and data are described in section 2, followed by SED fit analysis of the BSSs in section 3 and the characteristics of the hot companions in section 4. Finally, the discussion and concluding remarks on our work are presented in section 5 and 6.

\section{Observations and Data}
\subsection{UVIT Data}
UVIT is one of the payloads on the first Indian multi-wavelength space observatory \textit{AstroSat} launched on 28th September 2015, has been producing FUV and near-UV (NUV) images of good resolution \citep{Koshy2018, Mondal2018}.
The telescope has three channels: FUV (130-180 nm), NUV (200-300 nm) and visible (VIZ; 350-550 nm). The VIZ channel of the telescope operates mainly to correct the drift of the spacecraft. A detailed description of UVIT calibration and in-orbit performance are found in \cite{Subramaniam2016Calb}, \cite{TandonCalb2017} and \cite{Tandoninorbit}.
M67 observations were carried out in April 2017 (henceforth A17 data) and December 2018 (henceforth D18 data) using UVIT. We obtained near-simultaneous images in the FUV region with three filters during the first observation and with two filters during the second deeper observation. The journal of observation is given in Table \ref{tab:log}. We were unable to obtain NUV data due to a technical issue.
The raw images obtained from UVIT were reduced, stacked and corrected for drift to create science images using \textsc{ccdlab} \citep{Postma2017,Postma2020}. The UVIT images were analysed following the standard procedure using the {\sc daophot} package of {\sc iraf}. A point-spread function (PSF) photometry was performed on all filter images, and magnitudes were obtained after using the zero point magnitude for the respective filters and then corrected for aperture and saturation \citep{TandonCalb2017}. 

\begin{table}
    \centering
    \caption{Journal of observation}
    \label{tab:log}
    \begin{tabular}{cccc} 
    \hline
    Date & Filter & $\lambda_{eff}$$\pm$$\Delta\lambda$ & Exposure time\\ 
     & & (\AA)  & (s)\\
    \hline
    April 23. 2017&  F148W & 1481 $\pm$ 500 & 2261.16 \\ 
    April 23. 2017 & F154W & 1541 $\pm$ 380 & 2041.13\\
    April 23. 2017&  F169M  &1608 $\pm$ 290 & 2151.05\\
    December 19. 2018 & F148W & 1481 $\pm$ 500 & 6639.13\\
    December 19. 2018 & F169M & 1608 $\pm$ 290 & 6274.89\\
    \hline
    \end{tabular}
\end{table}

\begin{table*}
    \caption{Magnitudes of 6 BSSs detected with FUV filters of UVIT are listed along with their ID from \citet{Geller2015}, RA, Dec and optical magnitudes in V and B are taken from \citet{montgomery93}. The first three FUV magnitudes are from A17 data and, the remaining two FUV magnitudes are from D18 observations.}
    \label{tab:mag}
    \begin{tabular}{cccccccccc} 
    \hline
    WOCS ID & RA & DEC & V & B-V & F148W (A17) & F154W (A17) & F169M(A17) & F148W(D18) & F169M(D18)\\
     & deg & deg & mag & mag & AB mag & AB mag & AB mag & AB mag & AB mag\\ \hline
    1025 & 132.9071 & 11.6177 & 12.278 &0.385 & - & - & - & 19.596$\pm$0.032 & 19.145$\pm$0.05\\
    2013 & 132.9527 & 11.82116 & 10.92 & 0.31 & 16.549$\pm$0.028 & 16.405$\pm$0.028 & 16.187$\pm$0.005 &16.643$\pm$0.029 & 16.252$\pm$0.037\\
    3005 & 132.8858 & 11.8145 & 11.06 & 0.91 & 16.778$\pm$0.028 &16.575$\pm$0.019 & 16.448$\pm$0.036 & 16.825$\pm$0.018 & 16.442$\pm$0.022\\
    3013 & 132.7647 & 11.75095 & 11.32 & 0.295 & 17.889$\pm$0.044 &17.680$\pm$0.039&17.490$\pm$0.05 & 17.901$\pm$0.021& 17.496$\pm$0.021\\
    4006 & 132.8859& 11.84477& 12.257&0.26&17.728$\pm$0.038&17.660$\pm$0.032& 17.489$\pm$0.042& 17.751$\pm$0.024 & 17.384$\pm$0.026\\
    5005 & 132.8331& 11.78362& 12.126 & 0.458 & 19.893$\pm$0.089& 19.729$\pm$0.077&19.417$\pm$0.1&19.923$\pm$0.048 &19.468$\pm$0.043\\
    \hline
    \end{tabular}
 \end{table*}

\begin{figure}
 	\begin{center}
 	  \includegraphics[width=\columnwidth]{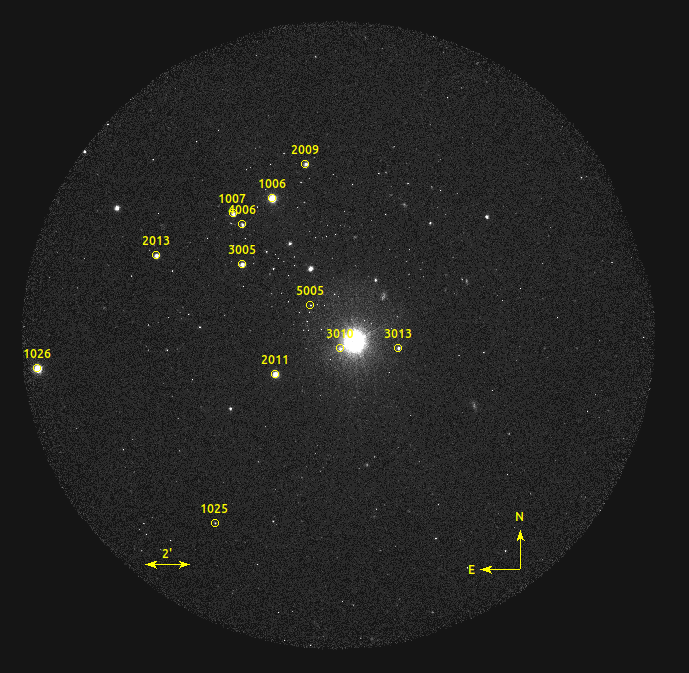}
     \end{center}
 	\caption{UVIT image of M67 obtained with F148W filter in December 2018 is shown. 12 BSSs detected with this filter are marked and labelled with their WOCS ID, the brightest star in the center of the image is saturated, and is a BSS.}
    \label{Fig: M67}
 \end{figure}

Among the 14 BSSs identified as members by \cite{Geller2015}, we have detected 11 BSSs in three FUV filter images from A17 data and 13 BSSs in two FUV filter images from D18 data. We have shown the UVIT image of FUV filter F148W obtained during the D18 run, and 12 BSSs are marked in Figure \ref{Fig: M67}. The hottest and brightest BSS is WOCS 1010, which is saturated in all the UVIT images. The study of 4 BSSs (WOCS 1006, WOCS 1007, WOCS 2011 and WOCS 2009) detected with UVIT was presented in \cite{Sindhu2019} and \cite{Jadhav2019}. As WOCS 3010 is located close to WOCS 8010 and the wings of the extremely saturated star WOCS 1010, we do not consider this star for further analysis. WOCS 1026 lies close to the edge of the image, and hence the photometry may not be reliable. We analyse the remaining 6 BSSs detected with UVIT in the present study. The magnitudes of 6 BSSs obtained from all the FUV filters are listed in Table \ref{tab:mag}.

\subsection{Archival data}
The archival data of UV to IR wavelength range was obtained from 0.9m Kitt Peak National Observatory (\citealt{montgomery93}; U, B, V, R, I), {\it Gaia} (\citealt{Gaia2018}; Gbp, G, Grp), 2MASS (\citealt{Cohen2003Calb2mass}; J, H, K$_s$) and {\it WISE} (\citealt{Wright2010wise}; W1, W2, W3). The log of observations of the archival data is listed in Table~1 of \cite{Sindhu2019}. We combine the archival data with the UVIT data to create a multi-wavelength SED spanning 0.14--11.5 $\mu$m range, after correcting for extinction in the respective bands \citep{Fitzpatrick1999, Indebetouw2005}. We have adopted a distance modulus V$-$M$_{v}$ = 9.6 $\pm$ 0.04 mag, Solar metallicity and E(B$-$V) = 0.041 $\pm$ 0.004 mag. 

\section{Analysis of Blue straggler stars}
The observed SEDs of 6 BSSs are constructed with multi-wavelength data, using Virtual Observatory SED Analyzer (VOSA; \citealt{Bayo2008}). 
The observed SEDs are fitted with a single spectrum covering the entire wavelength region using the Kurucz model (\citealp{Castelli1997} and updates).
The model is convolved with the filter responses to obtain the predicted synthetic flux for the corresponding temperature, surface gravity (log $g$) and metallicity values. The observed flux corrected for extinction is compared with the synthetic flux to get the best fit $\chi^2_{r}$ values amongst models, which must show minimal residual between the observed and synthetic flux to demonstrate a good fit.
The comparison of a single spectral fit in these 6 BSSs show significant deviation with respect to the expected synthetic flux and shows large $\chi^2_{r}$ values. Hence, we repeat the single spectral fit by ignoring observed flux points shorter than 1800 \AA \ in order to ignore the excess/deficient UV flux and refine the fit from NUV to the infrared (IR) wavelength band. This refit reveals any mismatch seen between the observed flux and model flux at the shorter wavelength and is used to detect any excess/deficient flux in the UV with respect to the expected UV flux for the BSSs. As the excess flux is observed only in the UV region of the SED, we fit the secondary component using Koester WD model spectra \citep{Koester2010}. We create composite models adding the expected flux from both the stars to fit the full observed SED from UV to IR. It is observed from literature studies \citep{Kepler2015} that most of the WD have a log $g$ of $\sim$8. Therefore, we have adopted a log $g$ of 7.75 for our fits, though the broadband fluxes used in the SED are insensitive to our choice.

\begin{figure*}
 	\begin{center}
     	
 	  \includegraphics[width=85mm, scale=2.0]{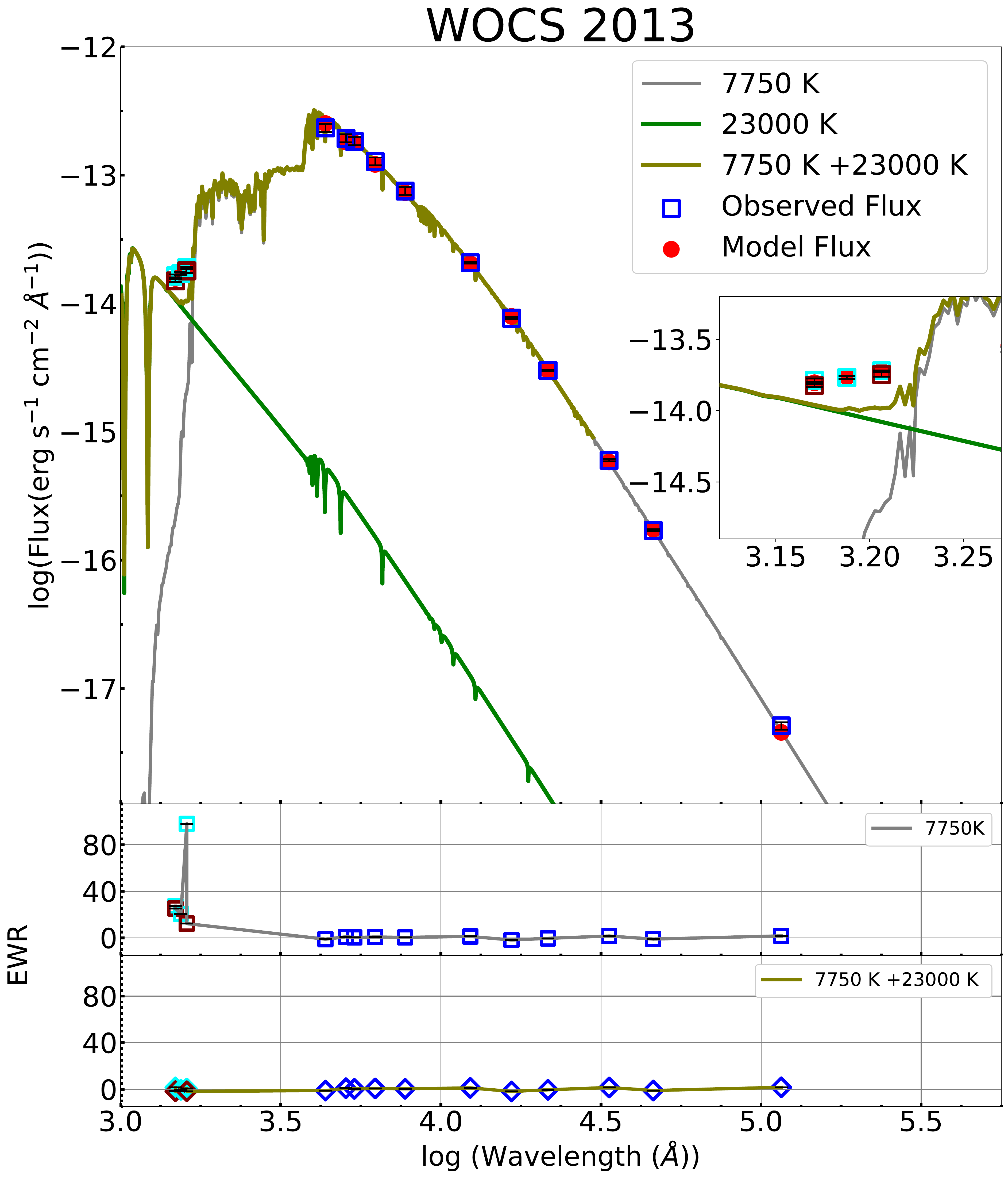}
 	   \includegraphics[width=85mm, scale=2.0]{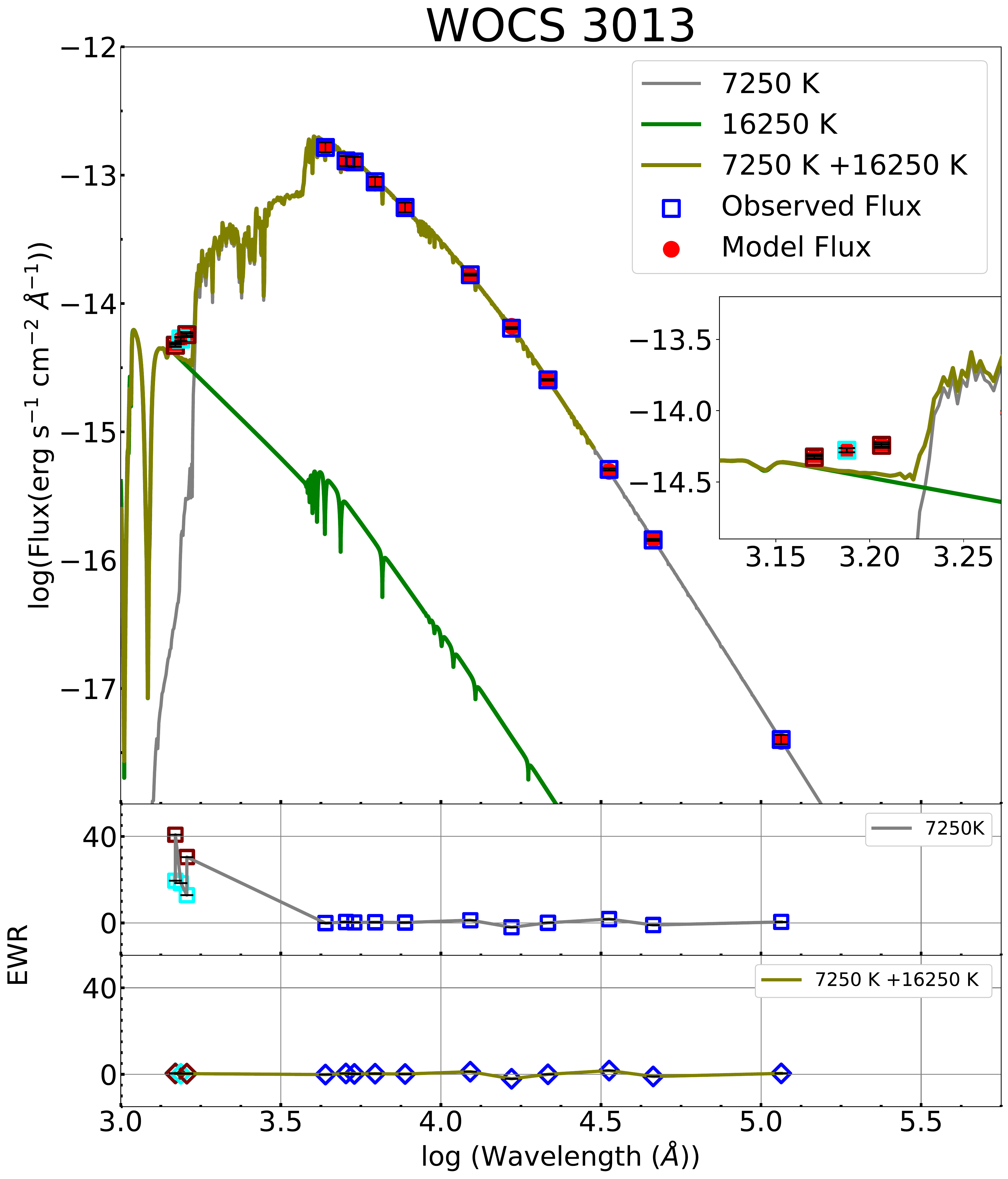}\\
 \includegraphics[width=85mm, scale=2.0]{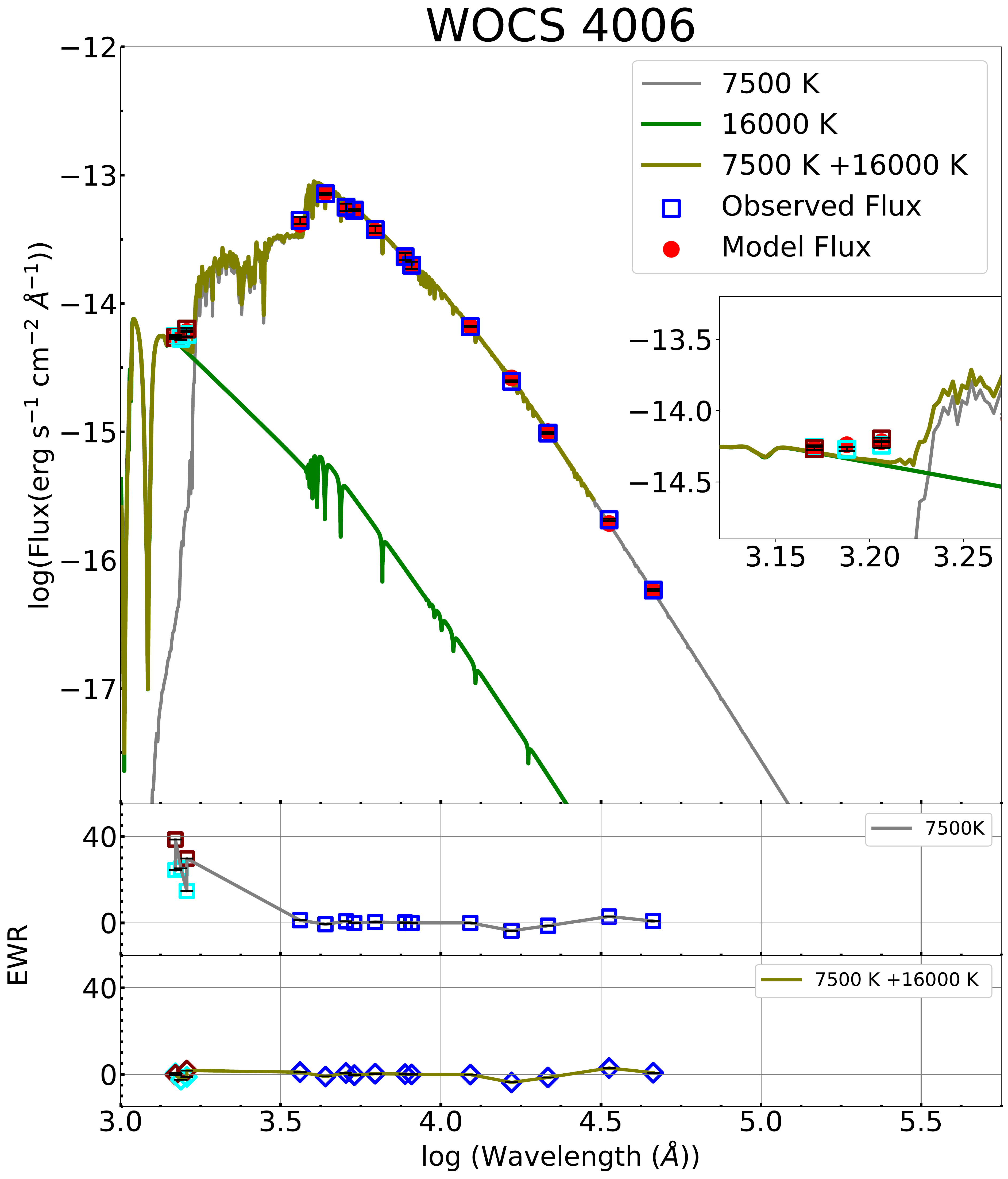}
 	   \includegraphics[width=85mm, scale=2.0]{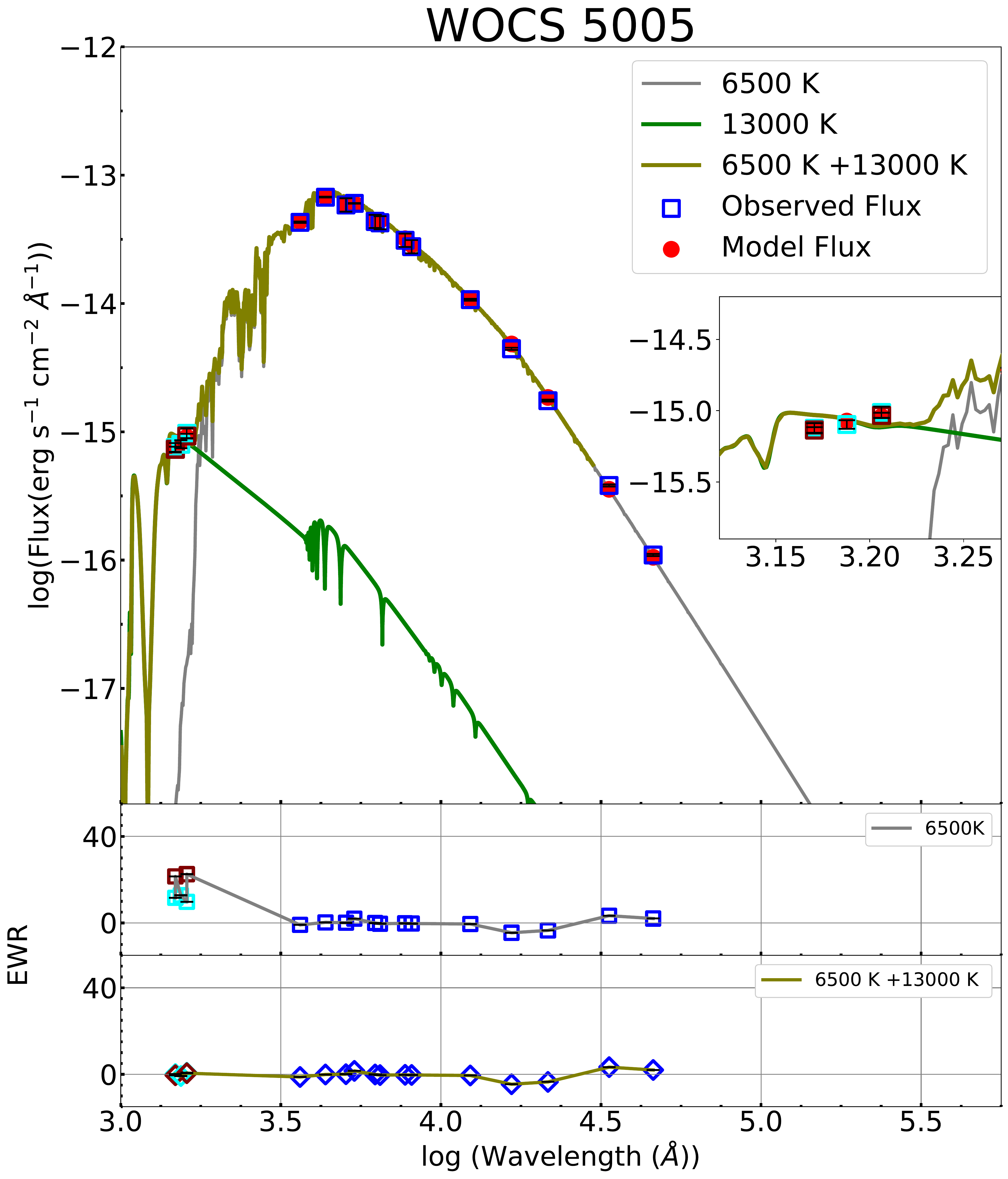}\\
 \end{center}
 \caption{SEDs of 4 BSSs are shown with an inset showing the UV region of the SED. Scaled and best fitting Kurucz spectrum (grey), Koester WD spectrum (green) and composite spectrum (olive) are shown, where the corresponding temperatures are shown in the panel. The observed photometric flux corrected for extinction from UV to IR region are shown as cyan (UVIT - A17), maroon (UVIT - D18) and blue squares (from optical to IR) with the corresponding composite synthetic flux in each band shown in red circle. The BSS identification number is shown in each figure. The EWRs of the SEDs are shown in the bottom panels for a single and composite fit. }
    \label{Fig: SED of 4BSS}
 \end{figure*}

\begin{figure*}
 	\begin{center}
 	  \includegraphics[width=85mm, scale=2.0]{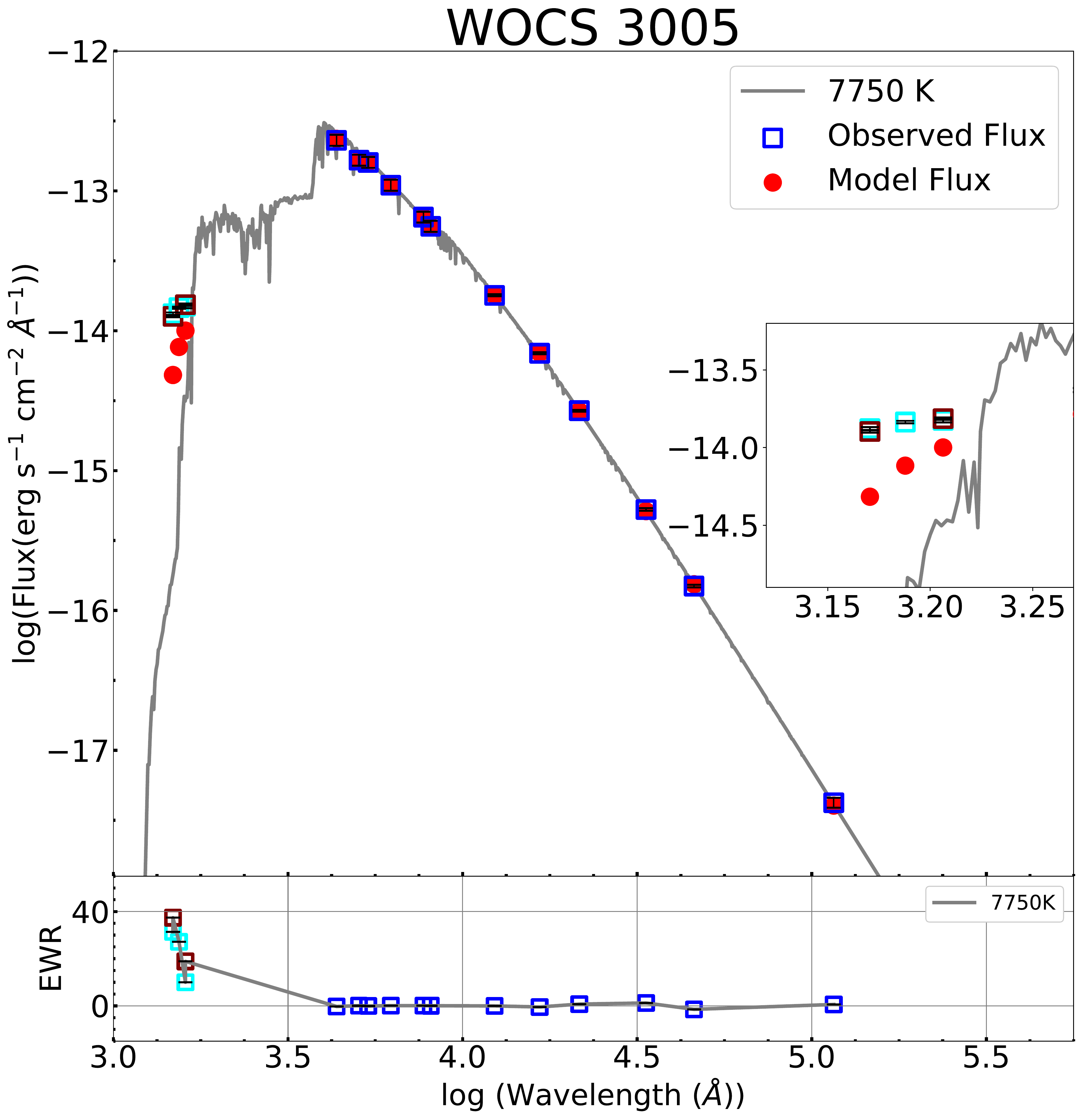}
 	   \includegraphics[width=85mm, scale=2.0]{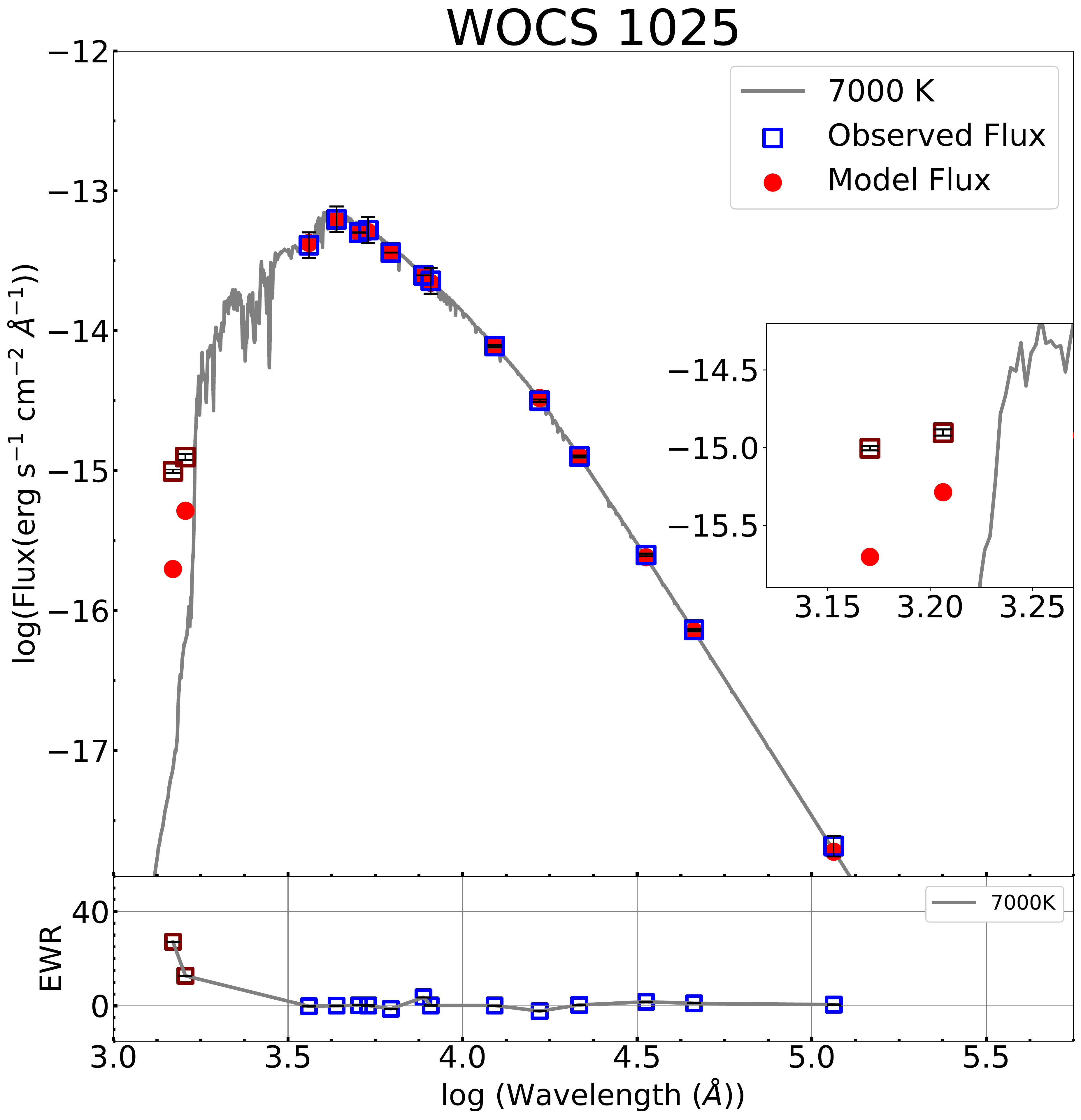}\\
 \end{center}
 	\caption{SEDs of 2 BSSs are shown with an inset showing the UV region of the SED. The scaled and best fitting Kurucz spectrum (grey) is shown, and the corresponding temperatures are shown in the panel. The observed photometric flux corrected for extinction from UV to IR region are shown as cyan (UVIT (A17)), maroon (UVIT (D18)), blue squares (from optical to IR), and the corresponding composite synthetic fluxes in each band are shown as filled red circles. The BSS identification number is shown in each figure. The residuals of the SEDs for a single fit are shown in the bottom panels.  
 	}
    \label{Fig: SED of 2BSS}
 \end{figure*}

We have shown the composite SED fits of 4 BSSs in figure \ref{Fig: SED of 4BSS} and single SED fits of two BSSs in figure \ref{Fig: SED of 2BSS}. The observed flux points after correcting for extinction from UV to IR wavelength region are shown. The WOCS ID of each BSS is shown at the top of each plot. UV region of the SEDs is zoomed and shown in the insets. The number of data points used for fitting the SEDs range from 15 to 18 with 3 fitting parameters, and the number of degrees of freedom (Ndof) ranges from 12 to 15; some BSSs have no detection in one or more filters. The error weighted residual (EWR) for both the fits (single and composite) are estimated using the formula:
\begin{equation}
EWR=\frac{Flux_{obs}-Flux_{model}}{Err_{obs}}
\end{equation}

The EWRs are plotted in the bottom panel of each SED fit of the BSSs, with the colour of the points and line connecting the flux points similar to the SED. The estimated fundamental parameters of the BSSs (A component) and their hotter companions (B component) obtained from the SED fits are listed in Table \ref{Tab:parameters} along with the $\chi^2_{r}$ values of the composite fits. We estimated the radius (R) of both the components using the scaling relation M$_{d}$ = $(R/D)^{2}$, where D is the distance to the cluster and M$_{d}$ is the scaling factor. 
We discuss below each of the BSSs, their SED fits and estimated parameters. 

\subsection{WOCS 2013 (S1267)}
WOCS 2013 is known to be SB1 with a period of 846 days and an eccentricity of 0.475 \citep{Latham1996}. The star has a rotational velocity of $vsini$= 60, 60.90 and 50 km s$^{-1}$ \citep{Pritchet1991,Motta2018,Latham1996}. \cite{Mathys1991} estimated the temperature (T$_{eff}$) and surface gravity to be 8010 K and log $g$= 3.91 using medium resolution spectroscopic observations. The star was also detected in the FUV image of \textit{UIT} \citep{Landsman98}. 

The SED of this BSS is fitted with a single spectrum to reveal an excess flux in the FUV, which is significant in all the UVIT filters. We used a double fit by including a hot component. The expected flux from a combination of two spectra with T$_{eff}$ = 7750 K and 23000 K fits the observed flux in all three/two UV filters of UVIT in two epochs. The values of the L/L$_{\odot}$ and R/R$_{\odot}$ are also small, suggesting the object could be a WD. The $\chi^2_{r}$ for the composite fit is much smaller than that for the single fit. We find that a WD like component fits the excess UV flux, suggesting a possible WD companion to this star. Thus, the hot sub-luminous component could be the binary companion to the BSS. 

\subsection{WOCS 3013 (S752)}
 \cite{Latham1996} derived the period to be 1003 days and having an eccentricity of 0.31 for WOCS 3013,  which is a SB1 with rotational velocity of $vsini$ = 60, 76.76 and 70 km s$^{-1}$ \citep{Pritchet1991, Motta2018, Latham1996}. 
\cite{Mathys1991} identified this star to be an Am star, and \cite{Sandquist2003} suggested this star to have a possible flare. The T$_{eff}$ and log $g$ of this star are 7640 K and 4.10 from medium resolution spectroscopic observations \citep{Mathys1991}. \cite{Landsman98} detected this star in the FUV image of \textit{UIT}.  
The SED of the WOCS 3013 is fitted well with a double fit; where the BSS is found to have a T$_{eff}$ = 7250 K and the hot component with a T$_{eff}$ = 16250 K for a log $g$ = 7.75. 
The radius of the hot component is 0.044 R$_{\odot}$ and sub-luminous, suggesting it to be a WD. 

\subsection{WOCS 4006 (S1280)}
WOCS 4006 is a $\delta$ scuti star with the highest number of frequency (41), and the second highest detected for any $\delta$ scuti star \citep{Bruntt2007}. The $\delta$ scuti pulsations in this star was first discovered by \cite{Gilliland1991}. They estimated log (L/L$_{\odot}$) = 0.90$\pm$0.07 and  T$_{eff}$ = 7244 K 
for this star. The fundamental radial mode of this oscillating star is identified to be 218 $\mu$m \citep{Gilliland1991, Gilliland1992}. The star is also an SB1 with rapid rotation of $vsini$ = 80, 100 km s$^{-1}$ \citep{Manteiga1989, Pritchet1991}. \cite{Mathys1991} determined  T$_{eff}$ = 8090 K and log $g$ = 4.33. 

The SED of this BSS is well fitted by a double SED, where the BSS is fitted with a T$_{eff}$ = 7500 K. The double fit brings down the $\chi^2_{r}$ value from 38.2 to 2.58. The hot component is found to have T$_{eff}$ = 16000 K and a radius of $\sim$0.051R$_{\odot}$. 
Our estimated log (L/L$_{\odot}$) for the BSS is found to be $\sim$ 0.86$\pm$0.021 and is consistent with the estimate of  \cite{Bruntt2007}. Though the BSS is known to be a binary, the orbital parameters are not known.

\subsection{WOCS 5005 (S997)}
WOCS 5005 is an SB1 with a rotation ($vsini$ $\leq$ 30, 20 km s$^{-1}$: \cite{Pritchet1991,Latham1996}). This star has a large period of 4913 days and an eccentricity of 0.342 \citep{Latham1996}. It was detected as an X-ray source by \cite{VandenBerg2004Chandra} in Chandra observation with an X-ray luminosity of 1.2$\times$10$^{29}$ erg s$^{-1}$. They suggested that the X-ray source is likely due to an undetected close binary. 

\begin{figure*}
 	\begin{center}
 	  \includegraphics[width=1.7\columnwidth, ]{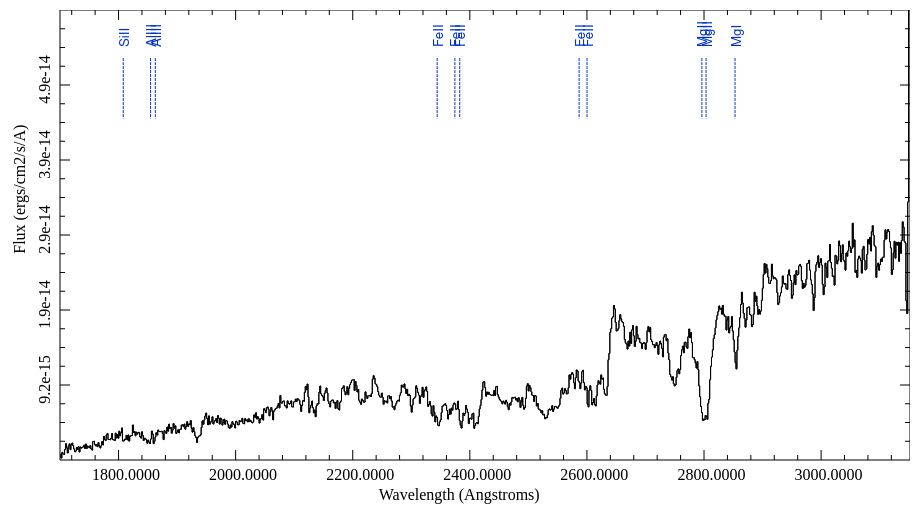}
 \end{center}
\caption{The HST spectrum of WOCS 5005 with G230L. The absorption profile of MgII {\it h+k} line is shown. 
 	}
    \label{Fig: HST of WOCS5005}
 \end{figure*}

The SED of this BSS is well-fitted by a double SED, where the BSS is fitted with a T$_{eff}$ = 6500 K and the excess UV flux is found to be significant. The double fit brings down the $\chi^2_{r}$ value from 55.2 to 3.6. The hot component is found to have T$_{eff}$ = 13000 K and a radius of $\sim$0.035R$_{\odot}$. This BSS was observed with the Space Telescope Imaging Spectrograph on {\it Hubble Space Telescope (HST)} at low spectral resolution on 21st April 2004, in 3 gratings (G230LB, G430L, G230L) and a prism with an exposure of 430 s, 24 s, 623 s and 41 s respectively. In Fig. \ref{Fig: HST of WOCS5005}, we have shown the spectrum obtained from HST G230L grating, where MgII {\it h+k} absorption line is observed. This feature is similar to those seen for 5 BSSs with \textit{International Ultraviolet Explorer} observations \citep{Sindhu2018}, ruling out chromospheric activity.

\begin{figure}
 	\begin{center}
 	  \includegraphics[width=85mm, scale=2.0]{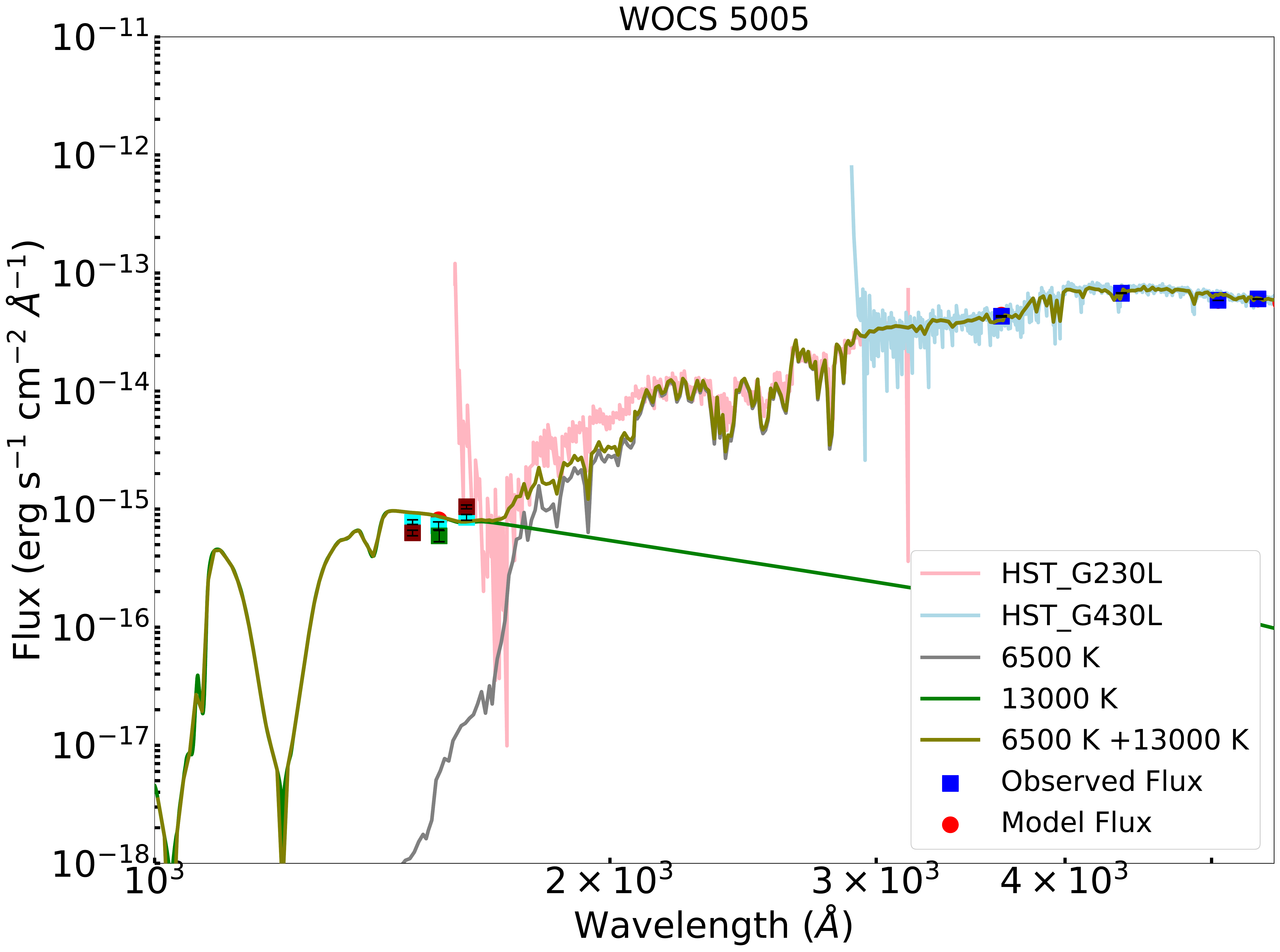}
\end{center}
\caption{The extinction corrected HST spectra of WOCS 5005 are shown in pink and blue, along with the observed fluxes shown as square points with colours similar to fig \ref{Fig: SED of 4BSS} and composite synthetic flux points are shown as red circles. The scaled and best-fitting theoretical spectrum of the Kurucz Model is plotted in grey, and the theoretical Koester spectrum (log $g$= 7.75) is shown in green. The composite spectrum for WOCS 5005 is shown in olive.}
    \label{Fig:Obs_HST}
 \end{figure}

In Fig. \ref{Fig:Obs_HST} we have shown the extinction corrected HST spectrum along with the observed photometric data points. The HST spectrum of G230L and G430L are shown in pink and blue, respectively. The data-set consisted of negative flux values, which are unphysical and have been removed. The FUV part of the spectrum is not fitted by the 6500 K Kurucz model spectrum (shown in grey), suggestive of FUV excess. The Koester model spectrum of T$_{eff}$ = 13000 K, shown in green, fits well with UV photometric points. The composite spectrum (shown in olive) fits the photometric points from UV to IR, which further confirms the presence of a cool and a hot component. 

\subsection{WOCS 1025 (S1195)}	
WOCS 1025 is an SB1 and identified as a binary member of the cluster based on radial velocity membership study \citep{Geller2015}. \cite{Latham1996} estimated the period and eccentricity of the BSS to be 1154 days and 0.066$\pm$0.082, respectively. They derived the rotational velocity of this star to be $vsini$ = 60 km s$^{-1}$.

The SED of this star is best fitted with a single spectrum of T$_{eff}$ = 7000 K to detect a significant UV excess. The observed flux is much higher when compared to the expected flux for the above temperature in two filters of the UVIT, as seen in figure \ref{Fig: SED of 2BSS}, as this star was detected in the second observing run with a slightly shifted UVIT pointing. We tried to fit a double SED for this BSS, but with just two UV points, we are unable to get a reliable T$_{eff}$ estimate for the hot component. If there is a hot secondary component, then it is likely to have T$_{eff}$ between 21,000 K and 50,000 K.
A circularised orbit and a large $v\ sini$ are indicative of recent MT episodes for this SB1. We suggest that this BSS has a probable hot companion, and more data in the UV are needed to estimate its parameters with confidence.

\subsection{WOCS 3005 (S1263)}	
\cite{Pritchet1991} derived the rotational velocity of WOCS 3005 to be $vsini$ $\leq$ 30 km s$^{-1}$. 
This star does not have a radial velocity variation, and is not a spectroscopic binary
\citep{Geller2015}. \cite{Sandquist2003} suggested this star to have a possible low-amplitude variations. \cite{Landsman98} detected this star in their \textit{UIT} observation of M67. 

The SED of this BSS is fitted with a single spectrum of T$_{eff}$ = 7750 K to detect a significant UV excess. The observed UV flux is much higher when compared to the expected flux for the above temperature in all three filters of the UVIT. We observed a higher flux residual in the F148W filter than other filters in the UV wavelength. 
Similar to WOCS 1025, we obtain a large range in T$_{eff}$ values for the hotter companion of this BSS, even though we have UV data from both our observing runs. As there is no indication of a binary companion for this BSS, further study is needed to understand the source of UV excess.

\begin{table*}
	\centering
	\caption{Fundamental parameters of the BSS  (A component) and WD companion (B component). The first and second columns list the WOCS and Sanders numbers, the third column identifies the component of the system, fourth, fifth, sixth and seventh columns list the T$_{eff}$, log $g$, Luminosity and Radius estimated for BSS \& WD companion respectively. $\chi^{2}_{r}$ for the composite fit is listed in the 8th column. }
	\vspace{0.2cm}
	\begin{tabular}{cccccccccc}
		\hline
			WOCS ID & S No.& Comp & T$_{eff}$ (K) & log $g$& L/L$_\odot$&R/R$_\odot$ & $\chi^{2}_{r}$\\[4.5pt] \hline

			2013 & 1267 & A & 7750$\pm$125 & 3.5 & 24.84$\pm$1.36 & 2.74$\pm$0.04 &  \\
                     	& & B & 23000$\pm$500& 7.75& 0.32$\pm$0.07& 0.036$\pm$0.004 & 1.78\\

\hline
		3013 & 752 & A & 7250$\pm$125 & 3.0 & 17.04$\pm$0.85 & 2.60$\pm$0.03 &  \\
	& & B & 16250$\pm$125& 7.75& 0.12$\pm$0.02 & 0.044$\pm$0.004 & 0.77\\
\hline
	4006 & 1280 & A & 7500$\pm$125 & 3.5 & 7.36$\pm$0.36 & 1.60$\pm$0.02 &  \\
	& & B & 16000$\pm$125&7.75 &0.15$\pm$0.03& 0.051$\pm$0.005 & 2.58\\		
\hline
	5005 & 997 & A & 6500$\pm$125 & 4.0 & 8.58$\pm$0.26 & 2.33$\pm$0.03 &  \\
	& & B & 13000$\pm$125& 7.75&0.032$\pm$0.006& 0.035$\pm$0.003 & 3.63  \\		
\hline		
1025 & 1195 & A & 7000$\pm$125 & 4.5 & 7.36$\pm$0.19 & 1.84$\pm$0.02 & \\
\hline
3005 & 1263 & A & 7750$\pm$125 & 3.0 & 21.6$\pm$1.0 & 2.58$\pm$0.03&\\
\hline
	\end{tabular}
	\label{Tab:parameters}
	\end{table*}

\section{Characteristics of the hot companions}
\begin{figure}
 	\begin{center}
 	    \includegraphics[width=85mm, scale=2.0]{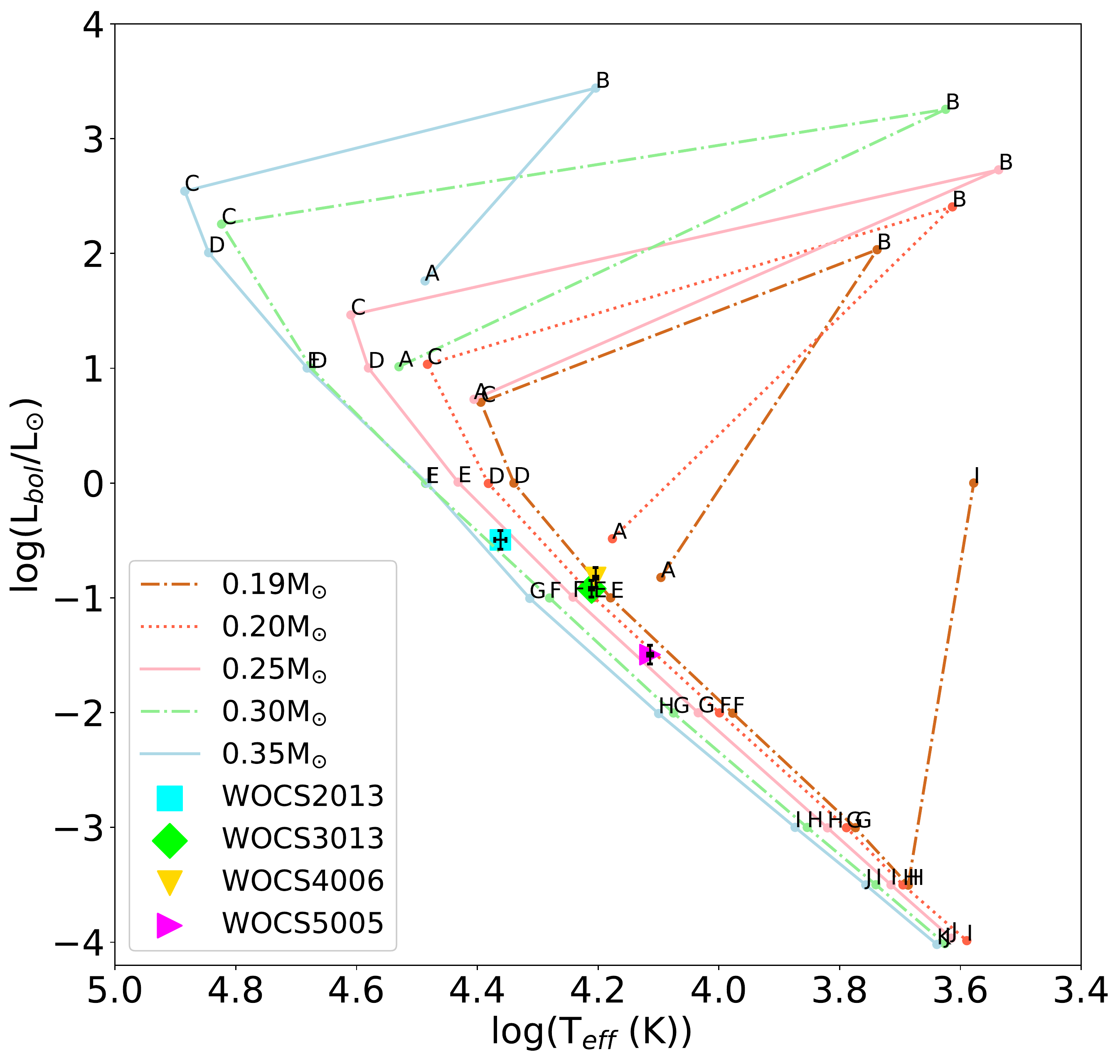}
    \end{center}
    \caption{HR diagram of WD companions to 4 BSSs (WOCS 2013-cyan square, WOCS 3013-light green diamond, WOCS 4006-yellow triangle, WOCS 5005-pink triangle) in M67 along the WD models from \citet{Panei2007} for masses 0.19, 0.2, 0.25, 0.3 and 0.35 M$_{\odot}$ are shown.}
    \label{Fig:HRD}
\end{figure}

The HR diagram of the hot companion of the BSSs is constructed using the parameters obtained from the binary SED fit and is shown in Fig. \ref{Fig:HRD} along with WD models taken from \cite{Panei2007} table 3. The model represents He-WD sequences with mass 0.1869, 0.2026, 0.2495, 0.3056 and 0.3515 M$_{\odot}$, with the characteristics of the models labelled by letters listed in their table 3. The evolution of each model begins at reference point A. At this stage, the MT is complete with the formation of He-WD (0 Myr). The end of the evolution is labelled as I, J and K depending on the He-WD mass. 

\subsection{Age and Mass}
In the HR diagram shown in Fig. \ref{Fig:HRD}, WOCS 2013 lies between the point E and F of 0.3 M$_{\odot}$ model. This reference point suggests that the low mass WD companion to WOCS 2013 was recently formed around $\sim$ 25 - 50 Myr. 
WOCS 3013 lies close to the reference point E of 0.2 M$_{\odot}$ model. This indicates that the low mass WD companion was formed $\sim$ 160 
Myrs ago. WOCS 4006 lies on the 0.19 M$_{\odot}$ model between D and E, this suggests that the MT completed around 190 
Myrs ago. WOCS 5005 is located on the 0.25 M$_{\odot}$ model between the reference point F and G, indicating that the low mass WD companion was formed between 40 - 300 Myr.
The masses of all the WD detected in this work are $<$ 0.35 M$_{\odot}$, which are low mass (LM) to extremely low mass (ELM) WD systems, with He core. We have listed the age and mass of the hot companion along with their kinematic properties from literature in table \ref{tab:Kinematics} for these stars.
\begin{table*}
\centering
\caption{Kinematic details of the BSSs are listed along with the age and mass of their WD companion.}
\label{tab:Kinematics}
\begin{tabular}{lccccc} 
\hline  
WOCS ID & Period (days)[L96]	&eccentricity [L96]& $v\ sin i$ (km s$^{-1}$) & WD age (Myr) & WD mass (M/M$_{\odot}$)		\\ \hline
WOCS 2013 & 846 & 0.475 & 60.90[M18] &25-50& 0.3\\ 
WOCS 3013 & 1003& 0.317 & 76.26[M18] &160& 0.2  \\
WOCS 4006 &     &     & 80.00[M89]  &$\sim$190& 0.19 \\
WOCS 5005 &4913 & 0.342 &20.00[L96]&40-300& 0.25 \\
WOCS 1025 & 1154 & 0.066 & 60.00[L96]& &\\
WOCS 3005 &     &      & 30.00[P91]& &\\
\hline
\multicolumn{6}{c}{M89: \cite{Manteiga1989}, L96: \cite{Latham1996}, M18: \cite{Motta2018}, P91: \cite{Pritchet1991}}\\
\hline
\end{tabular}
\end{table*}

\subsection{Period and Mass}
\begin{figure}
 	\begin{center}
 	  \includegraphics[width=\columnwidth]{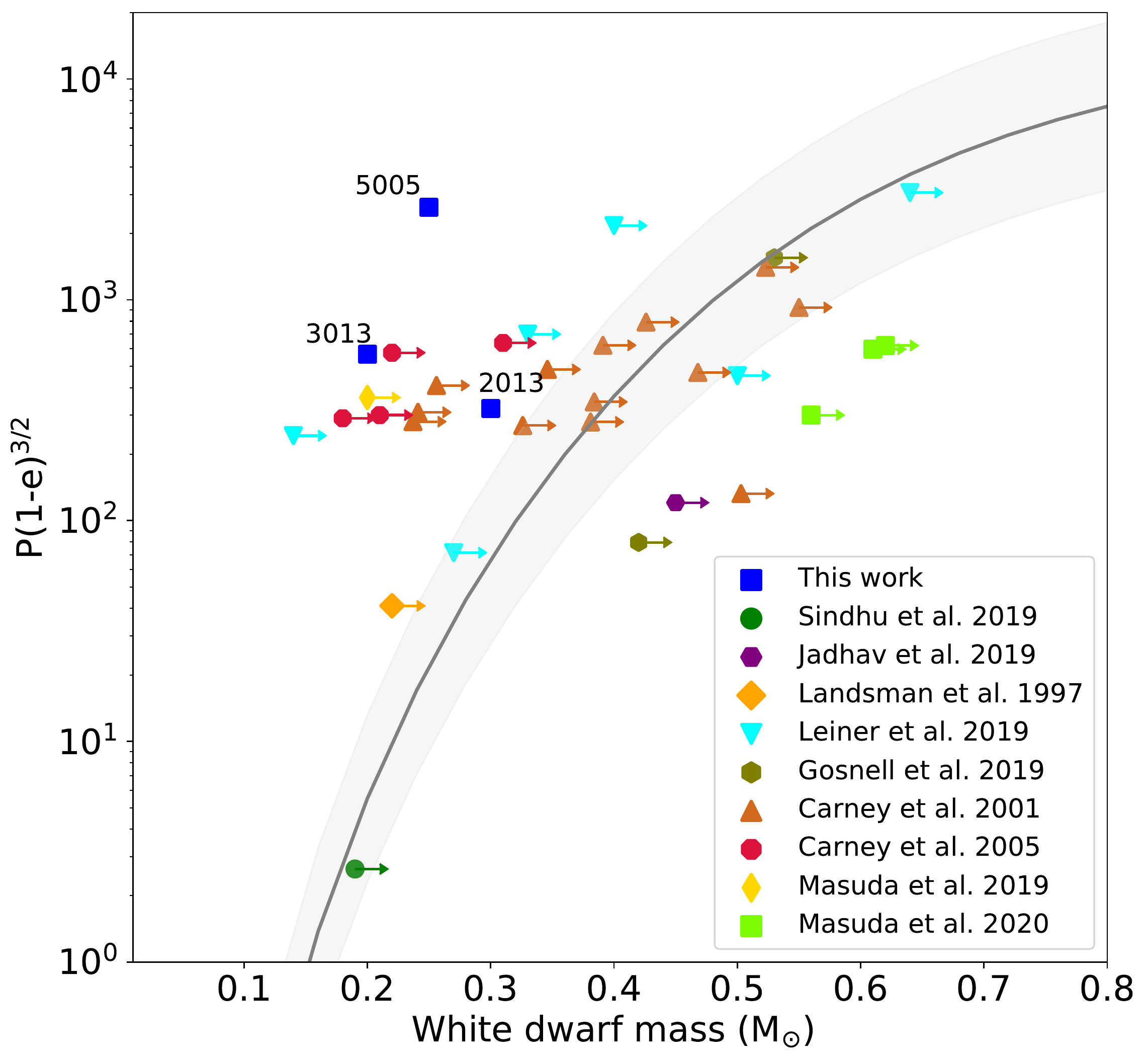}
 \end{center}
 	\caption{We have shown the masses of the WD binaries and their orbital separation at periastron for 3 BSSs (blue squares) of our sample. Green circle denote the WD binary to a BSS in M67 \citep{Sindhu2019}, purple hexagon and cyan triangles denote WD companions to blue lurkers in M67 \citep{Jadhav2019, Leiner2019}. WD companion to the YSS in M67 from \citet{Landsman1997} is denoted with orange diamond. Olive hexagon denotes the WD companion to the BSSs in NGC 188 \citep{Gosnell2015}. The sample of field BSS binaries from \citet{Carney2001,Carney2005} are denoted in orange triangles and red octagons. Self-lensing systems from Kepler are denoted in yellow diamond and light green square \citep{Masuda2019, Masuda2020}. The points shown here are lower limits of the secondary masses derived from binary mass function excluding those of our sample, \citet{Sindhu2019} and \citet{Jadhav2019}, which are obtained based on their location on the HR diagram and comparing to \citet{Panei2007} WD models. The grey line and region are from the theoretical orbital period--WD mass relationship from \citet{Rappaport1995} resulting from stable MT. WOCS 4006 with a WD companion is not shown as the orbital solution is not available.}
    \label{Fig:periastron_mass}
 \end{figure}

We show the binary orbital separations at periastron against the secondary mass for 3 BSSs of our sample, which have orbital solutions, in Figure \ref{Fig:periastron_mass}. Along with our sample, we have also shown other stars/BSSs with WD companions from literature. The figure shows the lower limit of the secondary mass obtained from the binary mass function, excluding those from this work, \cite{Sindhu2019} and \cite{Jadhav2019} which are obtained from their location in the HR-diagram and comparison to the \cite{Panei2007} He-WD models. In M67, a YSS and BSS detected with low mass WD companions are shown in this figure \citep{Landsman1997, Sindhu2019}.  Both of these stars are in agreement with the theoretical period-mass relation for stable MT for WD binary systems from \cite{Rappaport1995}, which is shown as a grey line in Fig. \ref{Fig:periastron_mass}, along with upper and lower limits of the relationship by a factor of 2.4 in the grey shaded region. This relationship is mainly applicable to systems with wide, nearly circular orbits with low mass WD companion. They have verified a modified (approximate) relation for non-negligible orbital eccentricity for wide binaries (Sirius and Procyon). They have found a good agreement between their theoretical results and observational data. We have taken the modified relation for our systems to check a similar agreement.

\cite{Jadhav2019} and \cite{Leiner2019} detected WD companions to blue lurkers in M67 indicated in the same figure. Blue lurkers are post-MT systems with masses lesser than the cluster turnoffs that were recently discovered in M67 \citep{Leiner2019}. Only a few blue lurkers from \cite{Leiner2019} seem to agree with the theoretical relation. WD companion to BSSs in NGC 188 detected by \cite{Gosnell2019} is shown, only one of the two WD binaries lie within the stable MT region of the figure. 
The sample of field BSS binaries from \cite{Carney2001,Carney2005} are plotted along with the self-lensing systems from Kepler  \citep{Masuda2019, Masuda2020}. All these stars have orbital periods $>$ 300 days and mostly lie outside the predicted parameter space for stable MT.  We find that two BSS in M67 and several field binaries are significantly outside the parameter space for stable MT. If the estimates for a stable MT are indeed correct, then these systems should have much shorter periods. Alternatively, the MT mechanism needs to be modified to include the formation of these systems.

\section{Discussion}
 The complex interaction of stars in a cluster dictates the evolution of single and binary stars. As M67 is a dynamically active cluster, it can produce exotic stellar systems such as BSSs \citep{Glebbeek2008}. 
 Among the BSSs observed in the cluster, about $\sim$79\% are in binaries, which are found to be in both short and long period orbit of varying eccentricities \citep{Latham1996, Sandquist2003}. Therefore, the high binary frequency and orbital period distribution suggests that the BSSs in an open cluster are formed through different channels. A chemical study of a few BSSs in M67 show no evidence of asymptotic giant branch MT as their surface abundances are similar to those of MS turn-off stars \citep{Motta2018, Shetrone2000}. Alternatively, measurement of rotation rates of BSSs can help in identifying their formation history, if they are products of a recent MT or collision/merger \citep{Leiner2019}. A recent MT is expected to leave behind a hot companion to the BSS (and blue lurkers), whereas mergers leave behind a single BSS. UV investigations have revealed the presence of hot WD companions to the BSSs and MS stars \citep{Knigge2008, Gosnell2015, Sindhu2019, Leiner2019, Jadhav2019, sahu_2019, Sindhu2020, Subramaniam2020}.

An excess in the UV flux can provide evidence for the presence of hot companions, as demonstrated by \cite{Sindhu2019} and \cite{Jadhav2019}, though older post-MT systems with cooler WD companions can escape detection, as do merger or collision products. In this study, we examine the SED of 6 more BSSs in M67 observed with UVIT and combine other multiwavelength photometric data from UV to IR bands. Combining all the photometric data of a star into an SED allows us to carry out quantitative analyses. As it is difficult to determine the nature of the companions with optical data alone, multi-wavelength SED is a widely used method to constrain the properties and nature of a companion. Though, we would not be able to determine the parameters such as metallicity, log{\it g} because broad-band SED fits are not sensitive to these parameters. 
Also, SED technique will not be able to de-convolve stars in case of triple systems, with two components having similar temperature.
It is also important to note that FUV flux measurements would be necessary to detect hot companions.

We could fit composite SEDs for 4 BSSs (WOCS 2013, WOCS 3013, WOCS 4006, WOCS 5005) and obtain the parameters for both the BSS and the hot companion. 
The binary SED fit suggests that the hot components for the 4 BSSs are LM/ELM WDs. We are unable to fit a double SED for the remaining 2 BSSs. WOCS 1025 has only two FUV photometric points, hence a reliable estimation was not possible. For WOCS 3005, we did observe UV excess in all the filters observed during both the epochs, but were unable to obtain the parameters of the secondary component, if present. WOCS 1025 is an SB1, and our estimates suggest a possible hot companion with T$_{eff}$ between 21,000 - 50,000K. WOCS 3005 is not a known binary, and the observed UV excess could be an intrinsic phenomenon arising from a chromospheric activity or flare.  
 
We constructed the HR diagram for the hot components of the 4 BSSs using the parameters obtained from the binary SED fits. 
Figure \ref{Fig:HRD} shows that the hot components of the 4 BSSs are comparable to He-WD of masses $\leq$ 0.35 M$_{\odot}$, which suggests them to be LM/ELM He WDs. The presence of LM/ELM WD as a secondary companion indicates that these objects are post-MT systems, as single stellar evolution restricts the formation of He-WD $\leq$ 0.4 M$_{\odot}$ within the Hubble time \citep{Brown2010, Istrate2016}.  We plotted a period--WD mass for the 3 BSSs which have orbital solutions and note that they do not fall in the region described by the theoretical-mass relation given by \cite{Rappaport1995} for stable MT. Therefore, a modified MT or some other mechanism is needed to explain the formation of these BSSs.

\cite{Melo2001} measured the $v\ sin i$ for a few stars MS, MS turn-off, end of MS turn-off and sub-giant stars, which have a rotational velocity less than 8 km s$^{-1}$. \cite{Motta2018} also estimated the $v\ sin i$ for a few MS turn-off of M67 and found them to be less than 20 km s$^{-1}$. WOCS 4006, WOCS 2013, WOCS 3013 and WOCS 1025 show rapid rotation, while WOCS 5005 and WOCS 3005 show relatively slow rotation $\leq$ 30 km s$^{-1}$. 
MT is one of the reasons suggested for the presence of fast rotation, as it transports angular momentum, resulting in a spin-up of the mass-accreting star \citep{Packet1981, DeMink2013, Matrozis2017}. Similarly, stellar collisions and mergers are also expected to yield rapidly rotating stars \citep{Sills2001, Sills2005}. 
Recently, \cite{Subramaniam2020} found a correlation between T$_{eff}$ of WD and $v\ sini$ of BSS. The BSSs with the hottest WD are fast rotators, whereas the slowest rotation is found for the BSS with the coolest WD (WOCS 5005). The cooling age estimates suggest that the WDs of 4006, 2013 and 3013 are formed recently (26 - 190 Myr), whereas the WD of 5005 has an upper age limit of $\sim$ 300 Myr. 
Thus, there is a good correlation here as well, which is suggestive of the spin-up of the BSSs due to MT. 

We detected 13 BSSs with three/two filters in the FUV using UVIT over two observing runs, of which 10 are analysed (this work,  \cite{Sindhu2019} and  \cite{Jadhav2019}. We detected 5 BSSs with LM/ELM WD companions, suggesting $\sim$50\% of the studied sample (5/10) and 35.7\% as the lower limit (5/14) for MT formation frequency of BSSs in M67. In the case of NGC 188, \cite{Gosnell2015} estimated MT frequency of $\sim$67\% among its BSSs.

 Case A MT scenario in M67 was studied by \cite{Tian2006}, but their models produced very few BSSs for the cluster, suggesting that other MT scenarios are required to produce BSSs in M67. 
 \cite{Lu2010} performed simulations and showed that Case B is as important as Case A in forming BSSs. They predicted that 5 BSSs could be formed through Case B MT and another three by dynamic merger processes. They also suggested that the BSSs formed via Case B MT are bluer and luminous than those produced by Case A. \cite{Sindhu2018} grouped the BSSs in M67 into three, of which group (a) (WOCS 1006, WOCS 2011, WOCS 1026, WOCS 3005, WOCS 1017, WOCS 2013, WOCS 1007) has the most luminous BSSs. The luminosity of WOCS 3013 estimated here makes it another member of the group (a). In total, we have characterised 6 of the 8 BSSs in this group, where LM/ELM WDs are found in 3 BSSs, with a possible hot companion in another BSS. As Case-B MT leads to ELM WDs, our study more or less matches with the \cite{Lu2010} predictions using mass transfer efficiency, $\beta$  = 1 models. They also showed that $\beta=0.5$ models produce lower luminosity BSS. WOCS 4006 and WOCS 5005 could be case-B MT products with low $\beta$ values. 
 
 It should be noted that binaries simulated by \cite{Lu2010} had periods of the order of tens of days, whereas the BSSs with detected LM/ELM WDs have much longer periods. This inconsistency is similar to what is seen in figure \ref{Fig:periastron_mass}, where only one BSS with ELM companion, WOCS 1007, is consistent with the stable MT models. Therefore, the question remains, how a Case-B MT is possible in these orbits with long periods. We list out a few possible explanations. First, M67 is known to be a cluster dominated by dynamics. N-body simulations of M67 by \cite{Hurley2005} found that the evolution and properties of two-thirds of the BSSs, including all found in binaries, have been altered by cluster dynamics, and nearly half would not have formed at all outside the cluster environment. It may be possible for the cluster dynamics to alter the orbits of these BS+WD systems resulting in longer eccentric orbits. It is also possible that these BS+WD are actually triple systems, where the estimated orbits are for the outer companion. The detected LM/ELM WD could be in a closer inner orbit that is yet to be characterised.  In this scenario, a hierarchical triple system could explain the  eccentricity of the outer orbit and the formation of a WD companion in a closer orbit. One of the reason that we are unable to decompose the SED for WOCS 3005 may be that it is a triple system. 
 The third possibility is that the stable MT mechanism needs modification to include these systems, as many are found not only in this cluster but also in the Galactic field.

 
 %
\section{Conclusion}

\begin{itemize}
    \item In this paper, we studied BSSs in the old open cluster M67 using UVIT images. The cluster was observed in three/two FUV filters (F148W, F154W and F169M) during two cycles (April 2017 and Dec 2018) of \textit{AstroSat}. We detected 13 BSSs, of which 6 BSSs are analysed in this work, 3 BSSs in \cite{Sindhu2019} and 1 BSS in \cite{Jadhav2019}. 
    \item All 6 BSSs showed excess flux in the UV, and we find 4 BSSs (WOCS 3013, WOCS 2013, WOCS 5005 and WOCS 4006) to fit well with a hot+cool composite model. The parameters of the hot companion have T$_{eff}$ in the range of 13000 K to 23000 K. 
    \item WOCS 1025 is likely to have a hot companion with T$_{eff}$ in the range, 21,000 - 50,000K. We detect excess UV flux in WOCS 3005, suggestive of a hot companion, but unable to characterise its properties.
    \item Comparing with \cite{Panei2007} models, the hot companions are found to have a mass range of 0.19 to 0.35M$_{\odot}$, suggesting them to be LM/ELM WDs. These BSSs are likely to be Case-B post-MT systems, as LM/ELM WD cannot be formed through single stellar evolution within the Hubble time. 
    \item We estimate a lower limit of 35.7\% for the MT pathway of BS formation in M67. 
    \item WOCS 2013, WOCS 3013 and WOCS 5005 are long-period binaries, hence challenges the formation through direct Case-B MT. To explain these systems, we propose that they could be dynamically affected to have long-period eccentric orbits, some may be in triple systems with the LM/ELM WDs in an unknown close inner orbit, or a modification in the MT mechanism is necessary.
\end{itemize}


\section*{Acknowledgements}
The UVIT project is a result of the collaboration between IIA, Bengaluru, IUCAA, Pune, TIFR, Mumbai, several centers of ISRO, and CSA. This publication makes use of VOSA, developed under the Spanish Virtual Observatory project supported by the Spanish MINECO through grant AyA2017-84089. The authors like to thank Prof. Ram Sagar for the useful discussion. SP acknowledges support from Council of Scientific \& Industrial Research (CSIR) for the SRF fellowship through grant 09/890(0005)/17 EMR-I. We thank the referee for the valuable inputs.

\section*{Data Availability}
The UVIT data can be accessed through ISRO AstroSat data archive depending  upon their proprietary period. 

\bibliographystyle{mnras}
\bibliography{ref.bib} 





\bsp	
\label{lastpage}
\end{document}